# Generating entangled Schrodinger cat states using a number state and a beam splitter

S.U. Shringarpure and J.D. Franson
Physics Dept., University of Maryland Baltimore County, Baltimore, MD 21250 USA

Passing a photon number state through a balanced beam splitter will produce an entangled state in which the phases of the two output beams are highly correlated. This entangled state can be viewed as a generalized form of a Schrodinger cat state where there is an equal probability amplitude for all possible phases. We show that Bell's inequality can be violated using this entangled state and two distant measuring devices that consist of a single-photon interferometer with a Kerr medium in one path, a set of single-photon detectors, and postselection based on a homodyne measurement. These entangled states are sensitive to photon loss and a violation of Bell's inequality requires either that the losses are inherently small or that their effects have been minimized using linear optics techniques [M. Micuda et al., Phys. Rev. Lett. **109**, 180503 (2012)]. Somewhat surprisingly, the use of the fair sampling assumption is not required for a violation of Bell's inequality despite the use of postselection if the measurements are made in the correct order.

## I. Introduction

Quantum mechanics violates Bell inequality, which rules out the possibility of local hidden-variable theories [1-5] as an alternative to quantum mechanics. The earliest experimental tests of Bell's inequality were based on entanglement between the polarizations or spins of two particles [6-12]. It was subsequently shown that Bell's inequality could be violated using continuous degrees of freedom, such as energy-time entanglement combined with two distant interferometers [13]. Here we note that a photon number state incident on a balanced beam splitter will produce an entangled state in which the phases of the two output beams are highly correlated [14,15]. This entangled state can be viewed as a generalized Schrodinger cat state where there is an equal probability amplitude for all phases.

We show that Bell's inequality can be violated using this entangled state and two distant measurement devices. Each of the measurement devices consists of a single-photon interferometer with a Kerr medium in one path, a set of single-photon detectors, and postselection based on a homodyne measurement. The use of postselection suggests that the fair sampling assumption may be required for a violation of Bell's inequality. Somewhat surprisingly, we show that the fair sampling assumption is not required if the measurements are performed in the correct order. Like other Schrodinger cats, these states are highly sensitive to photon loss. A violation of Bell's inequality requires that either the photon loss is inherently small or its effects have been minimized using linear optics techniques based on postselection [16].

It is well known that photon number states are highly nonclassical states of light [17] and that they are a useful resource for generating other kinds of nonclassical states. For example, a number state incident on a beam splitter has been used to herald an approximate cat state in one output mode by postselecting on the results of a homodyne measurement in the other output mode [18]. It has previously been shown that Bell's inequality can be violated using a variety of continuous variable states, homodyne measurements, or NOON states [19-22]. The approach described here is somewhat similar to earlier nonlocal interferometers [13,23], but the source of the entangled state is very different.

This paper is organized as follows. Section II outlines the basic approach. Section III derives the form of the entangled cat state at the output of the beam splitter. The nonlocal interference effects that can be observed using this entangled state are calculated in Section IV. Section V shows that Bell's inequality can be violated provided that the effects of photon loss are sufficiently small. In Section VI, we show that the fair sampling assumption is not required if the measurements are performed in the correct order. Section VII discusses an intuitive explanation for the origin of these effects, while Section VIII provides a summary and conclusions.

## II. Basic approach

We consider a situation in which a photon number state $|N\rangle$ is incident on a beam splitter with 50% transmission and reflection as illustrated in Fig. 1. As will be shown in the next section, the output state $|\psi\rangle$ from the beam splitter corresponds to a superposition of identical coherent states in each of the output beams. Since photon number and phase are conjugate variables, the phase of the input number state is totally uncertain and the output state corresponds to a superposition of all possible coherent-state phases between $0$ and $2\pi$. The nonlocal properties of this entangled state are the main focus of this paper.

We will show that the entangled state $|\psi\rangle$ can be used to violate Bell's inequality using two distant measurement devices as illustrated in Fig. 1. Each measurement device includes a single-photon interferometer (shown in red) with a Kerr medium in one of the two paths of the interferometer. The two output beams from the beam splitter also pass through the Kerr media, so that a phase shift will be applied depending on which path the single photons took. A fixed phase shift (not shown) is also included so that each of the beams will undergo a phase shift of $\pm\theta$. Single photon detectors $D_1$ through $D_4$ determine which path the single photons take when they leave the interferometers. Variable phase shifts $\sigma_1$ and $\sigma_2$ are also included in one path of the two interferometers as shown in Fig. 1.

Homodyne measurements are used to determine the quadratures $x_1$ and $x_2$ of the two beams after they have passed through the single-photon interferometers. We postselect on events in which the measured value of $x_1$ lies within a small range $\Delta x$ centered about some specific value $x_{1M}$, while $x_2$ lies in a range $\Delta x$ about $x_{2M}$. The combination of a single-photon interferometer with a Kerr medium in one path, the single-photon detectors, and postselection based on a homodyne measurement can be viewed as a compound measurement device. We will show that Bell's inequality can be violated in the usual way based on the output of the single-photon detectors $D_1$ through $D_4$ measured at four different settings of the parameters $\sigma_1$ and $\sigma_2$.

We will assume for the time being that the homodyne measurements are made after the single-photons have been detected in detectors $D_1$ through $D_4$, which simplifies the analysis. According to quantum mechanics, the same results would be obtained if the homodyne measurements were performed first. The advantages of the latter approach in ruling out hidden-variable theories will be discussed in Section VI.

The origin of these effects can be understood as being due to nonlocal quantum interference between two different probability amplitudes for obtaining quadrature measurements centered about $x_{1M}$ and $x_{2M}$. This will be described in more detail in the discussion of Section VII after we have calculated the properties of the system.

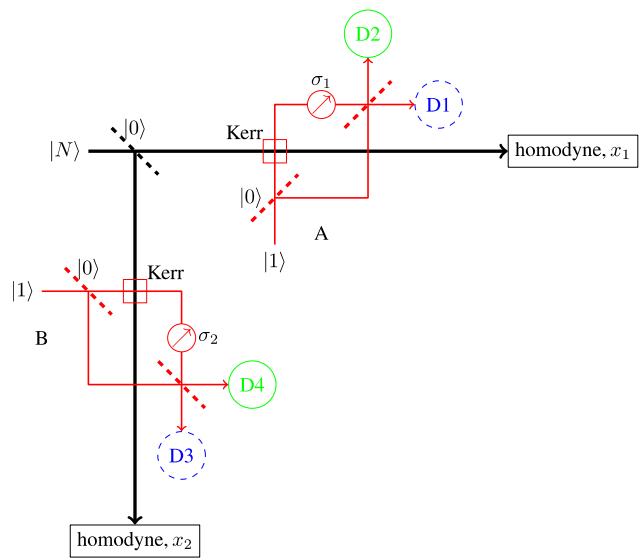

FIG. 1. A number state $|N\rangle$ incident on a balanced beam splitter will produce two output beams that are entangled in phase. A phase shift of $\pm\theta$ is applied to each of the beams using a pair of single-photon interferometers A and B with a Kerr medium located in one path, combined with a constant bias phase shift (not shown). Variable phase shifts $\sigma_1$ and $\sigma_2$ are introduced into one path of the single-photon interferometers and their outputs are measured using single-photon detectors $D_1$ through $D_4$. After the two beams have passed through the single-photon interferometers, their quadratures $x_1$ and $x_2$ are measured using homodyne detectors. Those events in which $x_1$ and $x_2$ lie within a small range $\Delta x$ centered about $x_{1M}$ and $x_{2M}$ are postselected. Bell's inequality can be violated in the usual way using the output of detectors $D_1$ through $D_4$ measured at four different settings of the parameters $\sigma_1$ and $\sigma_2$.

## III. Entangled state after the beam splitter

The effect of a balanced beam splitter can be described as usual by the unitary transformation

$$\hat{a}_1^\dagger \to \frac{\hat{a}_1^\dagger + i\hat{a}_2^\dagger}{\sqrt{2}} \tag{1}$$

and

$$\hat{a}_2^\dagger \to \frac{\hat{a}_2^\dagger + i\hat{a}_1^\dagger}{\sqrt{2}}. \tag{2}$$

Here $\hat{a}_1^\dagger$ and $\hat{a}_2^\dagger$ are the photon creation operators in the two input/output modes and we have used the common convention that the reflected component undergoes a phase shift of $\pi/2$.

The initial state $|\psi\rangle$ incident on the beam splitter is given by

$$|\psi\rangle = |N, 0\rangle, \tag{3}$$

where $|i, j\rangle$ will denote a state with $i$ photons in one mode and $j$ photons in the other mode. We will make use of the fact that a number state can be written as a superposition of coherent states [15]:

$$|N\rangle = \int_0^{2\pi} d\phi f_\phi |Re^{i\phi}\rangle. \tag{4}$$

Here $f_\phi$ is defined by

$$f_\phi \equiv \frac{e^{R^2/2} e^{-iN\phi} \sqrt{N!}}{2\pi R^N}, \tag{5}$$

and $|Re^{i\phi}\rangle$ denotes a coherent state with amplitude $R$ and phase $\phi$. $R$ is an arbitrary constant, but it will be convenient to choose the value $R = \sqrt{N}$.

Eq. (4) can be used to rewrite the initial state of the system before the beam splitter as

$$|\psi\rangle = \int_0^{2\pi} d\phi f_\phi |Re^{i\phi}, 0\rangle. \tag{6}$$

Here $|Re^{i\phi}, 0\rangle$ denotes a coherent state with amplitude $Re^{i\phi}$ in one input to the beam splitter and a coherent state with zero amplitude in the other input port.

It is well known that a coherent state incident on a beam splitter will produce a coherent state in the two output modes with amplitudes equal to the corresponding classical fields. As a result, the beam splitter transforms the state of the system in Eq. (6) into

$$|\psi\rangle = \int_0^{2\pi} d\phi f_\phi \left| \frac{R}{\sqrt{2}} e^{i\phi}, \frac{R}{\sqrt{2}} e^{i\phi} \right\rangle. \tag{7}$$

Here we have applied a phase shift of $-\pi/2$ in path 2 after the beam splitter to compensate for the factor of $i$ on reflection that appears in Eq. (1). (This phase shift is not shown in Fig. 1.)

The phase entanglement of the two beams is apparent in Eq. (7), which is qualitatively consistent with the results of Ref. [14] as well. All of the subsequent results can also be derived without using Eq. (4) by making use of the properties of the Hermite polynomials, as is described in the Appendix.

**IV. Nonlocal interference**

In this section, we will calculate the effects of the single-photon interferometers and show that there are two different probability amplitudes for obtaining homodyne measurement results of $x_{1M}$ and $x_{2M}$. Quantum interference between these two probability amplitudes can violate Bell's inequality. For the time being, we will only consider the situation in which single photons are detected in $D_2$ and $D_4$, which shows the nonlocal dependence on the phase shifts $\sigma_1$ and $\sigma_2$ in a straightforward way. In the following section, we will generalize the results to include photons detected in any of the four detectors $D_1$ through $D_4$, which can then be used to violate Bell's inequality in the usual way.

The single-photon interferometers inserted into paths 1 and 2 will be labelled by A and B, respectively. The state $|i, j\rangle_A$ will denote the case in which there are $i$ photons in the left path of interferometer A with $j$ photons in the right path, while $|i, j\rangle_B$ will denote the corresponding state in interferometer B. Including the single photons, the complete state of the system before the photons have entered the interferometers is given by

$$|\psi\rangle = \int_0^{2\pi} d\phi f_\phi \left| \frac{R}{\sqrt{2}} e^{i\phi}, \frac{R}{\sqrt{2}} e^{i\phi} \right\rangle |10\rangle_A |10\rangle_B. \tag{8}$$

After the single photons have entered their respective interferometers and passed through the first beam splitter, the state of the system becomes

$$|\psi\rangle = \int_0^{2\pi} d\phi f_\phi \left| \frac{R}{\sqrt{2}} e^{i\phi}, \frac{R}{\sqrt{2}} e^{i\phi} \right\rangle \\ \times \left( \frac{|10\rangle_A + i|01\rangle_A}{\sqrt{2}} \right) \left( \frac{|10\rangle_B + i|01\rangle_B}{\sqrt{2}} \right). \quad (9)$$

The presence of a single photon in the path with the Kerr media will produce a nonlinear phase shift and we assume that a constant phase shift is also applied so that the net phase shift is $\pm\theta$. As a result, the state of the system after the Kerr media can be written in the form

$$|\psi\rangle = \int_0^{2\pi} d\phi \frac{f_\phi}{2} \\ \times \left( |++\rangle_\phi |1010\rangle + i|+-\rangle_\phi |1001\rangle \\ + i|-+\rangle_\phi |0110\rangle + i^2 |--\rangle_\phi |0101\rangle \right). \quad (10)$$

Here we have introduced the notation

$$|+-\rangle_\phi \equiv \left| \frac{R}{\sqrt{2}} e^{i(\phi+\theta)}, \frac{R}{\sqrt{2}} e^{i(\phi-\theta)} \right\rangle, \quad (11)$$

with analogous definitions for $|-+\rangle_\phi$, $|++\rangle_\phi$, and $|--\rangle_\phi$. We have also used the more compact notation $|1010\rangle \equiv |10\rangle_A |10\rangle_B$, and so forth.

The single photons encounter the variable phase shifts $\sigma_1$ and $\sigma_2$ depending on which path they traverse as shown in Fig. 1. This transforms the state of Eq. (11) into

$$|\psi\rangle = \int_0^{2\pi} d\phi \frac{f_\phi}{2} \\ \times \left( e^{i(\sigma_1+\sigma_2)} |++\rangle_\phi |1010\rangle + i e^{i\sigma_1} |+-\rangle_\phi |1001\rangle \\ + i e^{i\sigma_2} |-+\rangle_\phi |0110\rangle + i^2 |--\rangle_\phi |0101\rangle \right). \quad (12)$$

Finally, the single photons exit the interferometers through another set of beam splitters which gives the state

$$|\psi\rangle = \int_0^{2\pi} d\phi \frac{f_\phi}{4} \\ \times \left( e^{i(\sigma_1+\sigma_2)} |++\rangle_\phi (|1010\rangle + i|1001\rangle + i|0110\rangle + i^2 |0101\rangle) \right. \\ + i e^{i\sigma_1} |+-\rangle_\phi (i|1010\rangle + |1001\rangle + i^2 |0110\rangle + i|0101\rangle) \\ + i e^{i\sigma_2} |-+\rangle_\phi (i|1010\rangle + i^2 |1001\rangle + |0110\rangle + i|0101\rangle) \\ \left. + i^2 |--\rangle_\phi (i^2 |1010\rangle + i|1001\rangle + i|0110\rangle + |0101\rangle) \right). \quad (13)$$

The case in which single photons are detected in $D_2$ and $D_4$ corresponds to the state $|0101\rangle$. Postselecting on that outcome gives the following unnormalized final state

$$|\psi\rangle = i^2 \int_0^{2\pi} d\phi \frac{f_\phi}{4} \\ \times \left( e^{i(\sigma_1+\sigma_2)} |++\rangle_\phi + e^{i\sigma_1} |+-\rangle_\phi + e^{i\sigma_2} |-+\rangle_\phi + |--\rangle_\phi \right). \quad (14)$$

The four terms in Eq. (14) correspond to the possible phase shifts in the two beams before they enter the homodyne detectors.

A single mode of the electromagnetic field is mathematically equivalent to a harmonic oscillator, and a homodyne measurement of the x-quadrature can be represented by the operator $\hat{x} = (\hat{a} + \hat{a}^\dagger)/\sqrt{2}$ with a suitable choice of units. As a result, it is convenient to use the position representation, where the usual wave function $\psi(x)$ is given by

$$\psi(x) = \langle x | \psi \rangle. \quad (15)$$

It can be shown [24] that the wave function $\psi_c(x)$ for a coherent state $|\alpha_0 e^{i\phi}\rangle$ of the field corresponds to a Gaussian wave packet of the form

$$\psi_c(x) = \frac{1}{\pi^{1/4}} e^{i p_0 x} e^{-(x-x_0)^2/2} e^{-i x_0 p_0 /2}. \quad (16)$$

Here $x_0 \equiv \sqrt{2} \alpha_0 \cos(\varphi)$ and $p_0 \equiv \sqrt{2} \alpha_0 \sin(\varphi)$. The overall phase factor of $e^{-i x_0 p_0 /2}$ is sometimes ignored, but it plays an important role [25] in superposition states such as in Eq. (9). In the coordinate representation, Eq. (14) gives

$$\psi(x_1, x_2) = \langle x_1, x_2 | \psi \rangle = \psi_{++}(x_1, x_2) + \psi_{--}(x_1, x_2) \\ + \psi_{+-}(x_1, x_2) + \psi_{--}(x_1, x_2), \quad (17)$$

where $\psi_{\pm\pm}(x_1, x_2)$ correspond to the four terms in Eq. (14). It will be convenient to choose the phase shift $\theta$ so that $N\theta = m(2\pi)$, where $m$ is an integer. In that case, Eqs. (13) and (16) can be used to show that

$$\psi_{++}(x_1, x_2) = -\frac{1}{4\sqrt{\pi}} e^{i(\sigma_1+\sigma_2)} \int_0^{2\pi} d\phi \, f_\phi e^{iR\sin(\phi+\theta)x_1} e^{iR\sin(\phi+\theta)x_2} \\ \times e^{-[x_1 - R\cos(\phi+\theta)]^2/2} e^{-[x_2 - R\cos(\phi+\theta)]^2/2} \left( e^{-i[R\sin(\phi+\theta)R\cos(\phi+\theta)]/2} \right)^2. \quad (18)$$

With $N\theta = m(2\pi)$, $\psi_{--} = \psi_{++}$ aside from the phase shift of $e^{i(\sigma_1+\sigma_2)}$. The cross-terms are given by

$$\psi_{+-}(x_1, x_2) = -\frac{1}{4\sqrt{\pi}} e^{i\sigma_1} \int_0^{2\pi} d\phi \, f_\phi e^{iR\sin(\phi+\theta)x_1} e^{iR\sin(\phi-\theta)x_2} \\ \times e^{-[x_1 - R\cos(\phi+\theta)]^2/2} e^{-[x_2 - R\cos(\phi-\theta)]^2/2} \\ \times e^{-i[R\sin(\phi+\theta)R\cos(\phi+\theta)]/2} e^{-i[R\sin(\phi-\theta)R\cos(\phi-\theta)]/2}, \quad (19)$$

with a similar expression for $\psi_{-+}$.

The probability $P(x_{1M}, x_{2M})$ of obtaining quadrature measurements that lie within a small interval $\Delta x$ about $x_1 = x_{1M}$ and $x_2 = x_{2M}$ is given by $P(x_{1M}, x_{2M}) = |\psi(x_{1M}, x_{2M})|^2 \Delta x^2$. (It is necessary to integrate over $x_1$ and $x_2$ for large values of $\Delta x$, as will be done in the following section.) It is possible to choose values of $x_{1M}$ and $x_{2M}$ such that $\psi_{+-}(x_{1M}, x_{2M})$ and $\psi_{-+}(x_{1M}, x_{2M})$ are negligible. In that case, Eq. (18) and the corresponding equation for $\psi_{--}(x_{1M}, x_{2M})$ can be used to show that

$$P(x_{1M}, x_{2M}) = \left| \left( e^{i(\sigma_1+\sigma_2)} \right) + 1 \right|^2 |\psi_{++}(x_{1M}, x_{2M})|^2 \Delta x^2 \\ = \gamma \cos^2[(\sigma_1 + \sigma_2)/2]. \quad (20)$$

Here $\gamma$ is a constant that depends on the choice of $x_{1M}$, $x_{2M}$, and $\Delta x$. The success rate for the postselection process (coincidence counting rate) depends on the value of $\gamma$ as will be discussed in the following section.

Eq. (20) shows that the coincidence measurements depend nonlocally on the sum of the phase shifts $\sigma_1 + \sigma_2$, which is characteristic of nonlocal interferometers such as that of Ref. [13]. A nonlocal interference pattern proportional to $\cos^2[(\sigma_1 + \sigma_2)/2]$ with a sufficiently high visibility indicates that Bell's inequality can be violated. This is shown to be the case in more detail in the following section.

### V. Violations of Bells inequality

The simple form of Eq. (20) depends on the assumption that the cross-terms $\psi_{+-}(x_{1M}, x_{2M})$ and $\psi_{-+}(x_{1M}, x_{2M})$ can be neglected. In order to investigate this possibility, the value of $|\psi_{++}(x_1, x_2)|^2 = |\psi_{--}(x_1, x_2)|^2$ is plotted as a function of $x_1$ and $x_2$ in Fig. 2(a). These results correspond to $N = 24$ and $\theta = \pi/4$, which satisfies the condition that $N\theta = m(2\pi)$. It can be seen that the phases of the two fields are highly correlated as expected. The magnitude squared of the wave function also shows an oscillatory behavior extending towards to the origin, which is due to the rapidly varying phase factor of $e^{-iN\phi}$ in the definition of $f_\phi$.

For comparison, Fig. 2(b) shows the magnitude squared of the cross-terms $|\psi_{+-}(x_1, x_2)|^2 = |\psi_{-+}(x_1, x_2)|^2$ as a function of $x_1$ and $x_2$. A phase shift of $\theta = \pi/4$ causes the phases of the two beams to become uncorrelated. In addition, the wave function is only appreciable inside a ring with a relatively narrow width. It can be seen that there are many choices of $x_{1M}$ and $x_{2M}$ where the cross-terms would be negligible compared to $|\psi_{++}(x_1, x_2)|^2$, which would give high-visibility nonlocal interference as described by Eq. (20). There are also regions where $|\psi_{++}(x_1, x_2)|^2$ is negligible compared to $|\psi_{+-}(x_1, x_2)|^2$, which would also allow high-visibility quantum interference between the $\psi_{+-}(x_{1M}, x_{2M})$ and $\psi_{-+}(x_{1M}, x_{2M})$ terms.

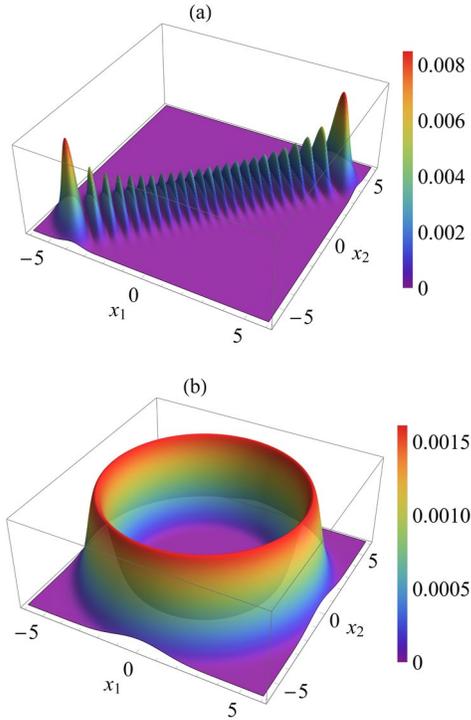

FIG. 2. Plots of the magnitude squared of the wave function in the coordinate representation as a function of $x_1$ and $x_2$. (a) Plot of $|\psi_{++}(x_1,x_2)|^2 = |\psi_{--}(x_1,x_2)|^2$. (b) Plot of the cross-terms $|\psi_{+-}(x_1,x_2)|^2 = |\psi_{-+}(x_1,x_2)|^2$. These results correspond to $N=24$, and $\theta = \pi/4$, which satisfies the condition that $N\theta = m(2\pi)$ where $m$ is an integer. (Dimensionless units.)

The normalized probability $P(x_{1M}, x_{2M})$ is shown in Fig. 3 as a function of the phase shift $\sigma_1$ in interferometer A for several values of the phase shift $\sigma_2$ in interferometer B. Here the postselected quadratures $x_{1M}$ and $x_{2M}$ were chosen to be $x_{1M} = x_{2M} = \sqrt{N}$, for which $\psi_{+-}(x_{1M}, x_{2M})$ and $\psi_{-+}(x_{1M}, x_{2M})$ are negligibly small. Bell's inequality [3] can be violated if the visibility of a nonlocal interference pattern of this kind is greater than $1/\sqrt{2}$ [26], which is the case for the results shown in Fig. 3.

There can be a significant contribution from the $\psi_{+-}(x_{1M}, x_{2M})$ and $\psi_{-+}(x_{1M}, x_{2M})$ terms for smaller values of $N$ or $\theta$. In that case, the interference pattern is no longer described by Eq. (20) and we must make use of the CHSH form of Bell's inequality introduced by Clauser, Horne, Shimony, and Holt [3]. We also need to generalize the results of the previous section to include the detection of single photons in any of the four detectors $D_1$ through $D_4$.

The effects of photon loss will be neglected initially but then included later in this section.

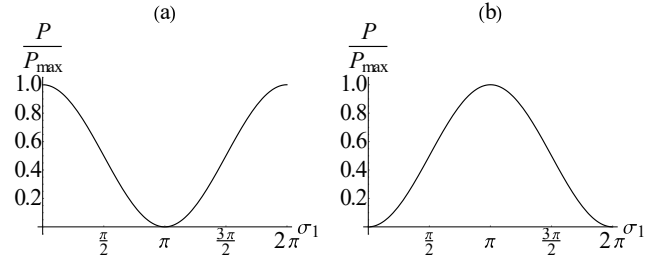

FIG. 3. The normalized probability $P/P_{max}$ of a postselected event for $x_{1M} = x_{2M} = \sqrt{N}$, plotted as a function of the phase shift $\sigma_1$ in interferometer A. (a) Phase shift $\sigma_2 = 0$ in interferometer B. (b) Phase shift $\sigma_2 = \pi$ in interferometer B. These nonlocal interference effects correspond to the same parameters as in Fig. 2 and they indicate that a violation of Bell's inequality should be possible.

The CHSH inequality requires two sets of measurement settings, which will be denoted by $\sigma_1 = \sigma_A$ or $\sigma'_A$ in interferometer A and $\sigma_2 = \sigma_B$ or $\sigma'_B$ in interferometer B. The result $a$ of the measurement obtained using $\sigma_1 = \sigma_A$ in interferometer A will be assigned the value $a = 1$ if a photon is detected in detector $D_1$, while it will be assigned the value $a = -1$ if a photon is detected in detector $D_2$. [3]. The results obtained in interferometer A using $\sigma'_A$ will be denoted $a' = \pm 1$ in a similar way, while the results obtained in interferometer B will be denoted $b = \pm 1$ or $b' = \pm 1$, depending on the choice of $\sigma_2$. The values of $\psi_{\pm\pm}(x_1, x_2)$ corresponding to the various single-photon detector outcomes can be calculated in the same way as in the previous section with the addition of a factor of $i$ when a single photon is reflected by a beam splitter.

The parameter $S$ in the CHSH form of the inequality is then defined as

$$S \equiv \langle ab \rangle + \langle a'b \rangle + \langle ab' \rangle - \langle a'b' \rangle. \quad (21)$$

Here $\langle ab \rangle$ denotes the average product of the measurement results $a$ and $b$, with a similar notation for the other three terms. The inequality $|S| \leq 2$ holds for all local hidden-variable theories.

In the example of interest here, the results are postselected on having obtained quadrature measurements of $x_1$ and $x_2$ within a range $\Delta x$ of $x_{1M}$ and $x_{2M}$. For small values of $\Delta x$, the properly normalized expectation values are therefore given by [3]

$$\langle ab \rangle = \frac{|\psi_{a=+1,b=+1}|^2 - |\psi_{a=+1,b=-1}|^2 - |\psi_{a=-1,b=+1}|^2 + |\psi_{a=-1,b=-1}|^2}{|\psi_{a=+1,b=+1}|^2 + |\psi_{a=+1,b=-1}|^2 + |\psi_{a=-1,b=+1}|^2 + |\psi_{a=-1,b=-1}|^2}, \quad (22)$$

with analogous results for the other expectation values. Here we have used the notation $\psi_{a=+1,b=+1} \equiv (\langle x_1, x_2 | \otimes {}_A\langle 10| \otimes {}_B\langle 10|)|\psi\rangle$ with a similar definition for the other three terms. The constant $\gamma$ and the factors of $\Delta x^2$ cancel out of these results.

We will first consider the case in which the range $\Delta x$ of accepted homodyne measurements is negligibly small. Fig. 4 shows a plot of $|S|$ as a function of $\sigma'_A$ and $\sigma'_B$, where the other measurement settings were held fixed at $\sigma_A = 0$ and $\sigma_B = \pi$. These results correspond to a relatively large photon number of $N = 24$ and $\theta = \pi/4$, as was used in Figs. 2 and 3. It can be seen that there are regions of the plot where $|S| > 2$ and Bell's inequality is violated, as would be expected from Fig. 3.

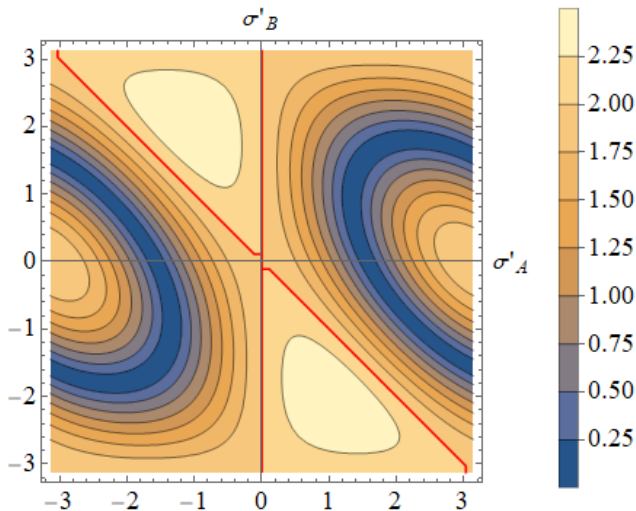

FIG. 4. A plot of the absolute value of the CHSH parameter $S$ as a function of the measurement settings $\sigma'_A$ and $\sigma'_B$. The other measurement settings $\sigma_A$ and $\sigma_B$ were held fixed at values of $0$ and $\pi$, respectively, while $N = 24$, $\theta = \pi/4$, and $x_{1M} = x_{2M} = \sqrt{N}$. The red contour lines correspond to $S = 2$, and it can be seen that there are large regions of the parameter space where the CHSH form of Bell's inequality is violated.

Fig. 5 shows a similar plot of $|S|$ for a more realistic value of $N = 4$. Although the interference pattern would no longer have the simple form shown in Eq. (20), it can be seen that there are still values of $\sigma'_A$ and $\sigma'_B$ where Bell's inequality can be violated.

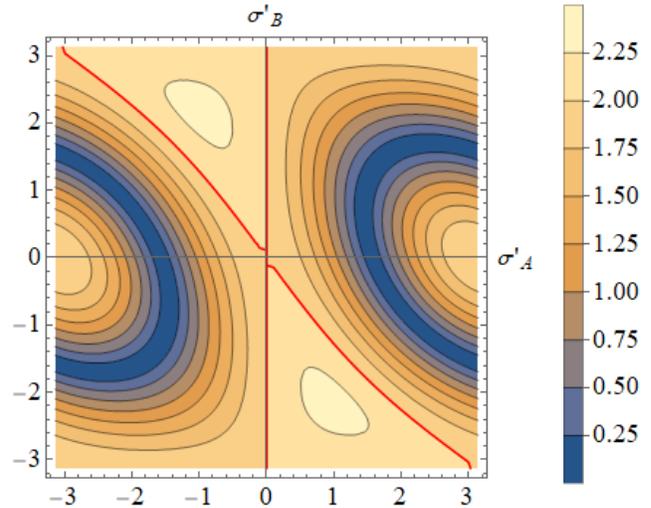

FIG. 5. Another plot of the absolute value of the CHSH parameter $S$ as a function of the measurement settings $\sigma'_A$ and $\sigma'_B$, where here the number of photons corresponds to $N = 4$. All of the other parameters are the same as in Fig. 4. It can be seen that there are still regions of the parameter space where $|S| > 2$ and the CHSH form of Bell's inequality is violated.

In order to obtain an acceptable counting rate in an experiment, it would be necessary to choose $\Delta x$ to be a significant fraction of the overall range of the homodyne measurement results, such as $\Delta x = 0.10\sqrt{N}$. This choice of $\Delta x$ can be shown to give a probability of success for the postselection process of 0.27% per pulse for $N = 4$, for example. With $\sigma_A = 0$, $\sigma'_A = 0.9$, $\sigma_B = \pi$, and $\sigma'_B = -2.2$, this value of $\Delta x$ gives a violation of Bell's inequality with $|S| = 2.3$. More generally, the maximum value of S is plotted in Fig. 6 as a function of $N$ for several values of $\Delta x$. Here $\sigma_A$, $\sigma'_A$, $\sigma_B$, and $\sigma'_B$ were varied to optimize $S$. These results were obtained by integrating the magnitude squared of the wave functions in Eq. (22) over the relevant range of $x_1$ and $x_2$.

It can be seen from Fig. 6 that Bell's inequality can be violated for all values of $N$, including $N = 1$. The value of $S$ is just above the hidden-variable limit of $S = 2$ for $N = 1$, and it gradually increases for larger values of $N$ up to the maximum value allowed by quantum mechanics of $2\sqrt{2}$ (the Tsirelson bound). It can also be seen that Bell's inequality can be violated for relatively large values of $\Delta x$, which would result in reasonable values for the probability of success.

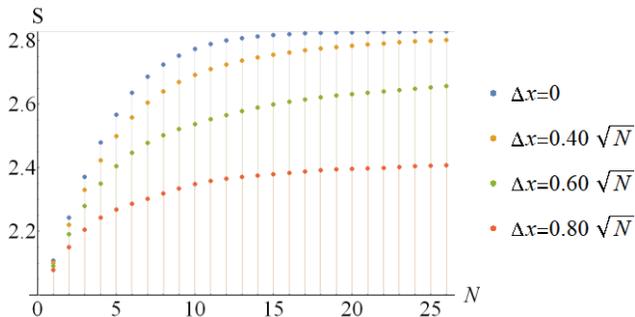

FIG. 6. Plots of the maximum value of the CHSH parameter $S$ as a function of the initial number $N$ of photons incident on the first beam splitter. The results are shown for several different values of the range $\Delta x$ of the homodyne measurement results that are accepted in the postselection process. These results neglect photon loss.

The results shown above neglect the effects of photon loss. Schrodinger cat states are very sensitive to photon loss, and the loss of even a single photon will typically produce a substantial amount of decoherence. The effects of photon loss during transmission were evaluated by including an additional beam splitter in both paths after the initial beam splitter of Fig. 1. The reflected components of the two beams provide which-path information regarding their phases, which reduces the amount of quantum interference [27]. The maximum value of the CHSH parameter S is plotted in Fig. 7 as a function of the mean photon loss $\bar{n}$ for several values of $N$. It can be seen that Bell's inequality can no longer be violated for $\bar{n}$ greater than $\sim 0.1$ photons for $N = 1$ and $\sim 0.2$ for $N = 4$, while larger values of $N$ allow a mean loss of $\sim 0.3$ photons. For a given channel loss, the value of $\bar{n}$ will be proportional to $N$, which favors the use of small values of $N$ in an experimental test.

Although the violation of Bell's inequality is sensitive to photon loss, Micuda et al. [16] have shown that an arbitrary quantum state can be transmitted through a lossy channel with negligible decoherence due to photon loss if the signal is noiselessly attenuated [16,28] before transmission, followed by noiseless amplification [29] after transmission. Roughly speaking, the noiseless attenuation can be used to reduce the intensity of the field to the point that the mean number of photons lost is much less than one and no significant which-path information is left in the environment. In principle, the phase-entangled states of interest here could violate Bell's inequality even after transmission through a lossy channel using techniques of that kind. Noiseless attenuation and amplification are both probabilistic, however, which results in an exponential reduction in the probability of success.

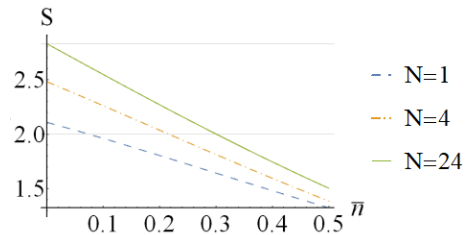

FIG. 7. Maximum value of the CHSH parameter $S$ as a function of the mean photon loss $\bar{n}$ for several values of the initial number $N$ of photons. These results assume that $\Delta x$ is negligibly small.

Another difficulty in an experimental test of Bell's inequality using this approach is the need to implement a Kerr phase shift at the single-photon level. Single-photon nonlinear phase shifts as large as $\pi/2$ have been demonstrated experimentally [30-37] but experiments of that kind remain challenging. Further improvements in those techniques would probably be required for the kind of experiments proposed here. Alternatively, a controlled phase shift (Kerr effect) can be implemented at the single-photon level using linear optics techniques [38,39], and we are currently investigating more efficient ways of doing that.

### VI. Fair sampling assumption

The postselection process based on the results of the homodyne measurements is required in order to violate Bell's inequality. As a result, one might suspect that the fair sampling assumption [40,41] may be needed. It will be shown in this section that the fair sampling assumption is not required provided that the homodyne measurements are completed before the phase shifts $\sigma_1$ and $\sigma_2$ are chosen at random. In that case, there is no opportunity for a hidden-variable model to bias the statistics as a result of the postselection process.

We have assumed up to now that the single-photon detection measurements in $D_1$ through $D_4$ were completed before the homodyne measurements in Fig. 1. That produces two possible phase shifts on beams 1 and 2 which give quantum interference effects in the subsequent homodyne measurements. But the results would be the same if the homodyne measurements were completed first, since the homodyne and single-photon detection measurements correspond to commuting operators. This can be accomplished by extending the length of the single-photon interferometers as illustrated in Fig. 8, which allows the settings of $\sigma_1$ and $\sigma_2$ to be chosen at random after the homodyne measurements have been completed.

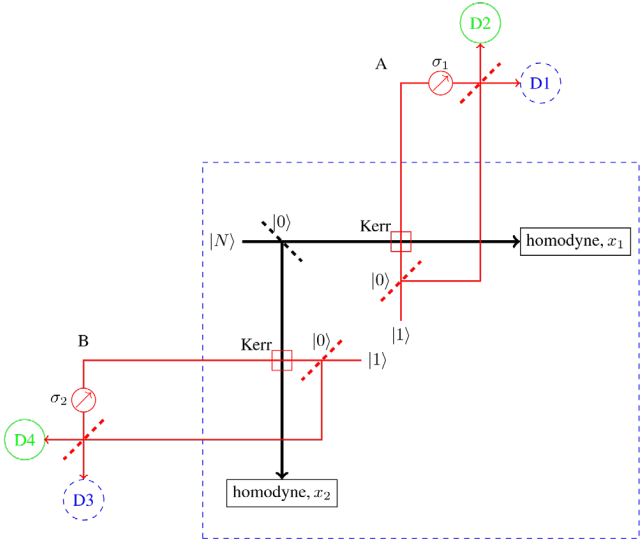

FIG. 8. Modification of the apparatus shown in Fig. 1 to avoid the need for the fair sampling assumption as a result of postselection. By extending the length of the interferometer arms, the homodyne measurement used for the postselection process can be completed before the phase settings $\sigma_1$ and $\sigma_2$ are chosen at random. Under those conditions, a local hidden-variable theory cannot bias the statistics from detectors $D_1'$ and $D_2'$. The combined system inside the dashed blue line can be viewed as a compound source that generates (heralds) the entangled states that are to be measured.

Fig. 9 compares the approach described here with a more conventional test of Bell's inequality using a pair of single-photon detectors with limited detection efficiency. Fig. 9(a) illustrates a conventional Bell's inequality experiment based on a pair of particles, such as two photons with entangled polarizations. In a local realistic theory, each photon is assumed to carry a set of hidden-variables $\{\lambda_i\}$ that are used to locally determine the outcome of two measurement devices with randomly-chosen settings $\theta_1$ and $\theta_2$, such as the orientation of two polarization analyzers. The two possible outcomes $a = \pm 1$ and $b = \pm 1$ of each measurement are then determined by single-photon detectors $D_{1a}$, $D_{1b}$, $D_{2a}$, and $D_{2b}$ as previously described. The measurement outcomes $a(\theta_1, \{\lambda_i\})$ and $b(\theta_2, \{\lambda_i\})$ are functions of $\theta_1$, $\theta_2$, and $\{\lambda_i\}$ in a hidden-variable theory. Bell's inequality can be derived in the usual way if the detectors have 100% detection efficiency. But for limited detection efficiencies, a photon entering a detector could have a probability of being detected that depends on the settings $\theta_1$ and $\theta_2$. That would bias the statistical results and allow a hidden-variable theory to violate Bell's inequality [40,41]. That possibility can be ruled out by using the fair sampling assumption or by using detectors with sufficiently high detection efficiencies.

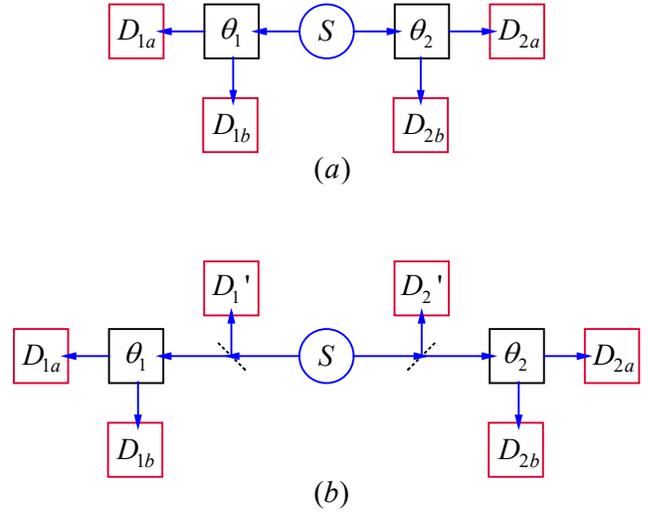

FIG. 9. Comparison of a conventional test of Bell's inequality with the approach described in the text. (a) A conventional Bell inequality experiment in which a pair of entangled particles are created by a source $S$ and propagate to two distant detectors with settings $\theta_1$ and $\theta_2$. The outcome of the measurements are recorded by detectors $D_{1a}$, $D_{1b}$, $D_{2a}$, and $D_{2b}$ with limited detection efficiencies. Given that a photon enters one of the detectors, the detection probability could depend on $\theta_1$ and $\theta_2$ in a hidden-variable theory, which would bias the statistics and require the use of the fair sampling assumption. (b) The experiment of interest here, where the results are postselected based on the measurement outcomes in detectors $D_1'$ and $D_2'$. The outcome of those measurements cannot depend on the choice of $\theta_1$ and $\theta_2$ if those measurements are completed before $\theta_1$ and $\theta_2$ are chosen at random. The fair sampling assumption would be required as usual if the detectors $D_{1a}$, $D_{1b}$, $D_{2a}$, and $D_{2b}$ have sufficiently low efficiencies.

For comparison, the approach of interest here is illustrated in Fig. 9(b). The optical pulses leaving the source contain an indeterminate number of photons along with a set of hidden-variables $\{\lambda_i\}$ that are used to locally determine the outcome of any measurements in a local hidden-variable model. A beam splitter separates part of the signal in each path and sends it to homodyne detectors $D_1'$ and $D_2'$, whose outcome will form the basis for the postselection process. The outcome of the postselection process would be determined by the hidden-variables $\{\lambda_i\}$ in a local hidden-variable theory.

After the homodyne measurements $D_1'$ and $D_2'$ have been completed, the settings $\theta_1$ and $\theta_2$ in the two measurement devices are chosen at random. In our example, $\theta_1$ and $\theta_2$ correspond to the phase shifts $\sigma_1$ and $\sigma_2$ in the extended-length single-photon interferometers shown in Fig. 8. In a local hidden-variable theory, the measurement outcomes $a(\theta_1,\{\lambda_i'\})$ and $b(\theta_2,\{\lambda_i'\})$ are determined by a new set of hidden-variables $\{\lambda_i'\}$ that are consistent with the outcome from the first set of measurements in $D_1'$ and $D_2'$. The probability distribution of these measurement outcomes can be viewed as conditional probabilities given the results obtained in $D_1'$ and $D_2'$. In any event, the new hidden-variables $\{\lambda_i'\}$ must determine the outcome of the subsequent measurements as recorded by detectors $D_{1a}$, $D_{1b}$, $D_{2a}$, and $D_{2b}$, in complete analogy with the role of the hidden-variables $\{\lambda_i\}$ in the conventional Bell's inequality test of Fig. 9(a).

Bell's inequality can then be proven as usual, based on the fact that the $\{\lambda_i'\}$ must determine the outcome of the subsequent measurements. The usual proof relies only on the requirement that the probability distributions associated with the $\{\lambda_i'\}$ must be normalized to unity (i.e., the hidden-variable theory must always produce an outcome) with all probabilities in the range of 0 to 1 (no negative probabilities allowed).

The fair sampling assumption is not required for the postselection process because the detection probabilities in $D_1'$ and $D_2'$ cannot depend on the subsequent choice of the settings $\sigma_1$ and $\sigma_2$. As a result, a hidden-variable model cannot take advantage of the limited detection efficiency of $D_1'$ and $D_2'$ to bias the statistics. That is not the case for the subsequent Bell-inequality measurement outcomes recorded by detectors $D_{1a}$, $D_{1b}$, $D_{2a}$, and $D_{2b}$, and the fair sampling assumption would still be required as usual if those detection efficiencies are sufficiently low.

The combined system consisting of the source S and homodyne detectors $D_1'$ and $D_2'$ can be viewed as an effective source that prepares an entangled state for a subsequent Bell inequality test. This is illustrated by the dashed-line box in Fig. 8. A compound source of this kind uses postselection to herald when an entangled state is ready to be measured, after which the hidden-variables $\{\lambda_i'\}$ must determine the outcomes of the measurements. From a conceptual point of view, this can be viewed as either a state preparation process or a preselection of quantum states, rather than postselection based on the results of the actual Bell-inequality test.

It can be seen from Eq. (7) that the postselection process does not create the entanglement between the phases in the two beams, which already exists before any measurements are made. The postselection provides a way to observe quantum interference between the probability amplitudes for states with different phases, which is required for a violation of Bell's inequality.

As a practical matter, the fair sampling assumption will be required in most experiments due to the limited detection efficiencies in $D_{1a}$, $D_{1b}$, $D_{2a}$, and $D_{2b}$. In that case, there is no need for the homodyne measurements to be space-like separated from the choice of $\sigma_1$ and $\sigma_2$, and the experimental apparatus of Fig. 1 would probably be easier to implement than the one in Fig. 8. The two experimental arrangements are equivalent according to quantum mechanics.

**VII. Discussion**

The quantum interference responsible for the violation of Bell's inequality can be understood in an intuitive way if $N \gg 1$. In that case, Eq. (7) describes a superposition of coherent states in each beam that is centered about a ring of radius $R = \sqrt{N}$ in phase space as illustrated in Fig. 10. The phases of the coherent states in the two beams are the same but totally uncertain, as illustrated by the red and blue circles.

A homodyne measurement on beam 1 will give a value of $x_1$ that corresponds to two possible values of the phase. For example, Fig. 11 illustrates the case where the measured value of $x_1$ is zero, which corresponds to a phase of $\pm\pi/2$. For simplicity, only the $\pi/2$ case is shown in the figure, where it is represented by a light blue circle. There are two ways of achieving a final phase of $\pi/2$. Both beams may have initially had a phase of $\pi/2-\theta$ followed by a phase shift of $\theta$ from the single-photon interferometers, which corresponds to the $\psi_{++}$ amplitude. Or both beams may have initially had a phase of $\pi/2+\theta$ followed by a phase shift of $-\theta$ from the single-photon interferometers, which corresponds to the $\psi_{--}$ amplitude. If the $\psi_{+-}$ or $\psi_{-+}$ amplitudes can be eliminated by the postselection process, then the total amplitude for this process to occur is $e^{i(\sigma_1+\sigma_2)}\psi_{++} + \psi_{--}$. Quantum interference between these two terms is responsible for the form of Eq. (20).

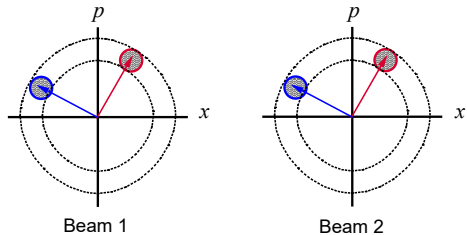

FIG. 10. Interpretation of the phase-entangled state in Eq. (7) in phase space, where $x$ and $p$ represent the position and momentum in the Wigner distribution. There is an equal probability amplitude for all possible phases at a distance of $R=\sqrt{N}$ from the origin. The small circles represent the uncertainty in the quadratures of the coherent states. A phase measurement in beam 1 will collapse the states in the two beams to approximate coherent states with equal phases [14]. Two possible results are illustrated by the two sets of arrows. (Dimensionless units.)

The $\psi_{+-}$ or $\psi_{-+}$ amplitudes correspond to the case where one beam undergoes a phase shift of $\theta$ while the other beam undergoes a phase shift of $-\theta$. For $N \gg 1$, those amplitudes do not overlap and they can be eliminated using postselection as illustrated by the dashed open circles in Fig. 11. But for small values of $N$, the radius $R$ will be reduced to the point that the uncertainty circles in Fig. 11 will overlap and the postselection process is ineffective. This is responsible for the reduction in the value of $S$ that can be seen in Fig. 6 for small values of $N$.

This interpretation is only approximately correct because of the uncertainty in the quadrates of a coherent state as well as the rapid variation in the factor of $\exp[-iN\varphi]$ in the integral of Eq. (7). Nevertheless, it does provide some insight into the origin of these effects and it was the initial motivation for our interest in the system shown in Fig. 1.

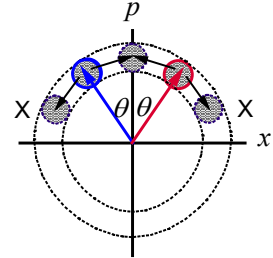

FIG. 11. Nonlocal quantum interference produced by applying a phase shift of $\pm\theta$ to the two output beams from the initial beam splitter shown in Fig. 1. This can be done using two single-photon interferometers containing a Kerr medium in one path. In this example, the results are postselected on obtaining a homodyne measurement of $x=0$ in both beams, which corresponds to a final phase of $\pm\pi/2$. (For simplicity, only the $\pi/2$ phase is shown.) One probability amplitude for this process to occur corresponds to an initial phase of $\pi/2+\theta$ in both beams, followed by a phase shift of $-\theta$ from the single-photon interferometers. A second probability amplitude corresponds to an initial phase of $\pi/2-\theta$ in both beams, followed by a phase shift of $\theta$ from the single-photon interferometers. Quantum interference between these two probability amplitudes can produce a violation of Bell's inequality. (Dimensionless units.)

### VIII. Summary and conclusions

A photon number state is one of the most basic examples of a nonclassical state of light. A number state incident on a balanced beam splitter will produce two output beams that correspond to a superposition of identical coherent states in each beam. The phase of the coherent states is totally uncertain but the same in both beams. An entangled state of this kind can be viewed as a generalized form of a Schrodinger cat state with an equal probability amplitude for all phases.

Bell's inequality can be violated using this entangled state and two distant measurement devices. Each measurement device consists of a single-photon interferometer with a Kerr medium in one path, a set of single-photon detectors, and postselection based on a homodyne measurement. The Kerr media produce a phase shift that depends on the path taken by the single photons through the interferometers. This gives two different probability amplitudes for obtaining the postselected value of the quadrature in the homodyne measurements. Nonlocal quantum interference between these probability amplitudes is responsible for the violation of Bell's inequality, which can occur for any number $N \geq 1$ of photons incident on the initial beam splitter.

Like other Schrodinger cats, these states are highly sensitive to photon loss. A violation of Bell's inequality requires that either the photon loss is inherently small or that its effects are minimized using linear optics techniques based on postselection [16]. The decrease in the CHSH parameter $S$ due to photon loss can be understood as being due to which-path phase information that is left in the environment.

The use of postselection suggests that the fair sampling assumption may be required for a violation of Bell's inequality. We have shown that the fair sampling assumption is not required if the homodyne measurements are performed before the parameters $\sigma_1$ and $\sigma_2$ in the Bell-inequality measurements are chosen at random. Quantum mechanics predicts the same results regardless of the order of the measurements.

Somewhat similar violations of Bell's inequality have previously been proposed using entangled Schrodinger cat states [23,27]. The main difference is that most entangled Schrodinger cat states only contain a superposition of two possible phases, whereas the entangled state produced by a number state and a beam splitter contains a continuous range of possible phases. This approach provides a relatively straightforward way to produce entangled cat states, especially for small values of $N$.

In principle, this technique could be used to distribute entangled pairs of photons in the output of the single-photon interferometers, provided that the photon loss is sufficiently small. More importantly, a number state is one of the simplest forms of a nonclassical state and the fact that it can be used in this way to violate Bell's inequality is of fundamental interest.

**Acknowledgements**

We would like to acknowledge valuable discussions with Todd Pittman. This work was supported in part by the National Science Foundation under grant number PHY-1802472.

**Appendix**

The results in the main text were derived using Eq. (4), which expresses a photon number state as a superposition of coherent states with all possible phases. In this appendix, we give an alternative derivation based on the properties of the Hermite polynomials. Although the results are equivalent to those in the text, they can be used to derive an analytic form for $\psi_{++}$ or $\psi_{--}$.

This initial state of the system with $N$ photons incident on the first beam splitter can be written in terms of the photon creation operators as

$$|\psi\rangle = \frac{\left(\hat{a}_1^\dagger\right)^N}{\sqrt{N!}}|0,0\rangle. \quad (A1)$$

Here the notation is the same as in the text. After passing through the beam splitter, Eq. (A1) can be used to write the transformed state as

$$|\psi\rangle = \frac{\left(\hat{a}_1^\dagger + i\hat{a}_2^\dagger\right)^N}{\sqrt{N!2^N}}|0,0\rangle. \quad (A2)$$

This can be rewritten using the binomial expansion as

$$|\psi\rangle = \sum_{n=0}^{N} i^n \sqrt{\frac{{}^N C_n}{2^N}} |N-n,n\rangle \quad (A3)$$

where ${}^N C_n$ are the binomial coefficients.

We can then simplify each of the terms in Eq. (14) by integrating over the phase $\phi$. The results are that

$$|\psi_k\rangle = -p_k \frac{1}{4} \sum_{n=0}^{N} q_k^n \sqrt{\frac{{}^N C_n}{2^N}} |N-n,n\rangle, \quad (A4)$$

where $p_1 = e^{i(\sigma_1+\sigma_2+N\theta)}$, $p_2 = e^{i(\sigma_1-N\theta)}$, $p_3 = e^{i(\sigma_2+N\theta)}$, $p_4 = e^{-N\theta}$, $q_1 = q_4 = 1$ and $q_2 = q_3^* = e^{i2\theta}$. The index $k$ corresponds to each of the four terms in Eq. (17).

The dimensionless position-basis representation of the final state is given by the inner product $\langle x_1, x_2 | \psi \rangle = \psi(x_1, x_2)$. We can use the position-basis representation of the number states,

$$\langle x|n\rangle = \sqrt{\frac{e^{-x^2}}{n!2^n\sqrt{\pi}}} H_n(x), \quad (A5)$$

to obtain

$$\psi_k(x_1,x_2) = -p_k \frac{e^{-(x_1^2+x_2^2)/2}}{4\sqrt{\pi N!}} \sum_{n=0}^{N} q_k^n \frac{{}^N C_n}{2^N} H_{(N-n)}(x_1) H_n(x_2). \quad (A6)$$

Here $H_n(x)$ is the $n^{th}$ Hermite polynomial.

We can further simplify Eq. (A6) for $k=1$ and $k=4$ to the analytical form

$$\psi_k(x_1,x_2) = -p_k \frac{e^{-(x_1^2+x_2^2)/2}}{4\sqrt{2^N \pi N!}} H_N\left(\frac{x_1+x_2}{\sqrt{2}}\right). \quad (A7)$$

When the cross-terms terms corresponding to $k=2$ and $k=3$ are negligible, Eq. (A7) gives an interference pattern that is equivalent to Eq. (20) in the text, but without any complicated integrals over $\phi$.